\documentclass[11pt]{article}
\pdfoutput=1 

\usepackage{jcappub}
\usepackage{amsfonts,amssymb,amsmath,verbatim,mathtools,graphicx,bm,booktabs,tabularx,multirow,relsize,framed,orcidlink}
\usepackage{xcolor}
\usepackage[T1]{fontenc}
\usepackage[utf8]{inputenc}
\usepackage{subfig}
\usepackage{tikz-cd, tikz}
\usepackage{graphicx}
\usepackage{float}
\usepackage{wrapfig}
\usepackage{algorithm}
\usepackage{listings}
\usepackage[noend]{algpseudocode}
\usepackage{appendix}
\usepackage{listings}

\usepackage{nicefrac}
\usepackage{relsize}
\usepackage{bbold}
\usepackage{revsymb}

\newcommand{\unige}{D\'epartement de Physique Th\'eorique, Universit\'e de Gen\`eve, 24 quai Ernest Ansermet, 1211 Gen\`eve 4, Switzerland}

\newcommand{\gwsc}{Gravitational Wave Science Center (GWSC), Universit\'e de Gen\`eve, CH-1211 Geneva, Switzerland}

\title{\boldmath Analytical results for the power-law sensitivity curve of stochastic gravitational-wave backgrounds}

\author[1,2]{Enis Belgacem\,\orcidlink{0000-0003-4920-0911}}
\affiliation[1]{\unige}
\affiliation[2]{\gwsc}

\emailAdd{Enis.Belgacem@unige.ch}

\abstract{We derive analytically some general features of the power-law sensitivity curve. They include an exact parametric equation, a formula for the peak sensitivity and a proof of convexity in log-log plot. A few conceptual points are also clarified.}

\begin{document}
\maketitle
\flushbottom

\section{Introduction}\label{intro}

The detection of a stochastic gravitational-wave background (SGWB) is part of the large ongoing effort put by the gravitational wave (GW) science community into the new window of observation opened since the first direct GW detection \cite{LIGOScientific:2016aoc}. SGWBs are expected to be produced by both cosmological and astrophysical processes. Among the proposed cosmological origins there are first-order phase transitions, cosmic strings and the amplification of primordial quantum fluctuations of the gravitational field either during inflation or in alternative scenarios. Astrophysical SGWBs stem from the incoherent superposition of GWs from many independent and unresolved individual astrophysical sources, like compact binary coalescences (see~\cite{Belgacem:2024ohp} for a thorough derivation of the corresponding spectral density), supermassive black hole binaries, core-collapse supernovae, pulsars and magnetars. A treatment of these sources can be found in~\cite{Maggiore:2018sht, Christensen:2018iqi} and references therein. Evidence for a SGWB consistent with a population of supermassive black hole binaries has been claimed in~\cite{NANOGrav:2023gor}. The next generation of ground-based interferometers like Einstein Telescope \cite{Hild:2008ng, Punturo:2010zz, Hild:2010id, ET:2019dnz, Branchesi:2023mws} and Cosmic Explorer \cite{Reitze:2019iox, Evans:2021gyd, Evans:2023euw}, especially when combined in a network, will be characterized by a remarkable sensitivity that will allow to resolve a huge fraction of individual compact binary coalescences. The detection of a background of cosmological origin is then conceivable; see~\cite{Belgacem:2024ntv} for a first-principle treatment of the effect of the residual astrophysical background (given by the unresolved events and the reconstruction error on the resolved ones) on the detectability of a cosmological background.

The sensitivity of a network of detectors to a stationary, Gaussian, unpolarized and isotropic SGWB can be characterized by the Power-Law integrated Sensitivity (PLS) curve , introduced in~\cite{Thrane:2013oya}. In this paper we want to make an analytical study of the properties of the PLS curve. Along the way, we will also clarify some technical points that have been implicitly assumed in~\cite{Thrane:2013oya}. The paper is organized as follows. In section \ref{sec:defs} we write the basic equations and definitions. In section \ref{sec:PLSmain} we recall the primary definition of the PLS curve as the envelope of a family of backgrounds, we give a simple interpretation and we derive an exact parametric equation for it. In section~\ref{sec:PLSisfunc} we show that the PLS curve is also a function and we also justify, with details in appendix~\ref{app:eq30_details}, the form of this function assumed in~\cite{Thrane:2013oya}. In section~\ref{sec:PLSshape} we derive some other general properties on the shape of the PLS (function, by then) and an exact formula for the peak sensitivity. Finally, in section~\ref{sec:concl} we summarize the main findings and draw the conclusions.

\section{Definitions and basic equations}\label{sec:defs}

A SGWB is a stochastic process consisting of a superposition of GWs with all possible frequencies, directions of propagation and polarizations. In the linearized gravity approximation of General Relativity and working in the transverse-traceless (TT) gauge, a realization of the SGWB at time $t$ and position $\vec{\bf{x}}$ is
\begin{equation}
\label{eq: h_ijTT}
h_{ij}^{\rm TT}(t,\vec{\bf{x}})=\int_{-\infty}^{\infty}df\int_{S^2}d^2\hat{\bf{n}}\,
e^{-2\pi i f(t-\hat{\bf{n}}\cdot\vec{\bf{x}} /c)}\sum_{A\in\{+,\times\}}\tilde{h}_A(f,\hat{\bf{n}})\, e^A_{ij}(\hat{\bf{n}})\, ,
\end{equation}
where $e^A_{ij}(\hat{\bf{n}})$ are the components of the two polarization tensors (with polarization labeled by an index $A\in\{+,\times\}$ and with spatial indices $i,j\in\{1,2,3\}$), $S^2$ is the unit 2-sphere and $\tilde{h}_A(f,\hat{\bf{n}})$ are the Fourier modes of the SGWB for each polarization $A$ at a given frequency $f$ and direction of propagation specified by the unit vector $\hat{\bf{n}}$. The speed of propagation of GWs is equal to the speed of light $c$. Negative frequencies have been included in the decomposition~(\ref{eq: h_ijTT}), with the condition that
\begin{equation}
\label{eq: reality hij to Fourier}
\tilde{h}_A(-f,\hat{\bf{n}})=\tilde{h}_A^*(f,\hat{\bf{n}})\,,
\end{equation}
so that $h_{ij}^{\rm TT}(t,\vec{\bf{x}})$ is real, as it must be since it is a metric perturbation.

We will denote by $\langle\dots\rangle$ a stochastic ensemble average. For a stationary SGWB, the expectation value $\langle h_{ij}^{\rm TT}(t,\vec{\bf{x}})\rangle$ is constant in time and it can actually be assumed to vanish because any constant value does not contribute to GWs. Then also for the Fourier components of each polarization we can assume that $\langle\tilde{h}_A(f,\hat{\bf{n}})\rangle=0$.
If the stationary SGWB is also Gaussian then it is uniquely characterized by its two-point functions. Assuming also that the SGWB is unpolarized and isotropic, then\footnote{For astrophysical backgrounds, the corrections to eq.~(\ref{eq: two-pt GW}) due to a finite observation time, as well as the anisotropies and polarization induced by shot noise, have been derived in\cite{Belgacem:2024ohp}.}

\begin{equation}
\label{eq: two-pt GW}
\langle\tilde{h}_A^*(f,\hat{\bf{n}})\tilde{h}_{A'}(f',\hat{\bf{n}}')\rangle=\frac12 S_h(f)\,\delta_{A A'}\,\delta(f-f')\,\frac{\delta^2(\hat{\bf{n}},\hat{\bf{n}}')}{4\pi}\,,    
\end{equation}
where $\delta^2(\hat{\bf{n}},\hat{\bf{n}}')$ is the Dirac delta on the unit 2-sphere and $S_h(f)$ is the single-sided spectral density of the background. $S_h(f)$ is real, positive and $S_h(-f)=S_h(f)$.
When studying the energy content of a SGWB it is useful to define the dimensionless quantity
\begin{equation}
\Omega_{\rm GW}(f)=\frac{1}{\rho_c}\frac{d\rho_{\rm GW}}{d\ln f}\,,
\end{equation}
where $d\rho_{\rm GW}/d\ln f$ is the energy of the SGWB per unit spatial volume and per unit logarithmic frequency interval (integrated over all directions of propagation), while $\rho_c$ is the current value of the critical energy density to close the Universe, given by $\rho_c=3c^2H_0^2/(8\pi G)$, where $H_0$ is the Hubble constant and $G$ is Newton's constant. The relation with the spectral density is (see e.g. eq.~(7.202) of~\cite{Maggiore:2007ulw})
\begin{equation}
\label{eq:OmegaGW_Sh}
\Omega_{\rm GW}(f)=\frac{4\pi^2}{3H_0^2}f^3S_h(f)\,.
\end{equation}

Let us consider a network of $N_{\rm det}\geq2$ detectors, labeled by an index $a=1,2,\dots,N_{\rm det}$. If the noise $n_a(t)$ is stationary then the expectation value\footnote{For the average over different realizations of the noise, we are using the same notation $\langle\dots\rangle$ that we adopted for the SGWB, even if these two averages have a different meaning and physical origin. To be more precise we can actually define $\langle\dots\rangle$ to be the average over {\it both} these two statistical ensembles. In this way, when considering for example $\langle n_a(t)\rangle$ or $\langle\tilde{n}_a(f)\rangle$, the average over the realizations of the SGWB has a trivial action as it plays no role. Similarly, for quantities like $\langle\tilde{h}_A(f,\hat{\bf{n}})\rangle$ the average over noise realizations is irrelevant.} $\langle n_a(t)\rangle$ is time-independent and, without loss of generality, it can be assumed that $\langle n_a(t)\rangle=0$. Then, going to frequency domain via a Fourier transform,
\begin{equation}
\tilde{n}_a(f)=\int_{-\infty}^{+\infty}dt\,n_a(t)\,e^{2\pi i f t}\,,
\end{equation}
we also have $\langle\tilde{n}_a(f)\rangle=0$. If the noise is also Gaussian and there are no correlations between noise in different detectors, then the statistical properties are completely specified by the (single-sided) noise spectral density $S_n^{(a)}(f)$ of each detector, with
\begin{equation}
\label{eq: noise uncorr}
\langle\tilde{n}^*_a(f)\tilde{n}_b(f')\rangle=\frac12 \delta_{ab}\,S_n^{(a)}(f)\,\delta(f-f')\,.
\end{equation}
$S_n^{(a)}(f)$ is real, positive and $S_n^{(a)}(-f)=S_n^{(a)}(f)$.

We consider a cross-correlation search for a SGWB at the given network of detectors, assuming the same common time of observation $T$ for all the pairs of detectors. In a common positive frequency band $[f_{\rm min}, f_{\rm max}]$ the optimal total signal-to-noise ratio (SNR or $S/N$) is (\cite{Allen:1997ad,Maggiore:2007ulw})
\begin{equation}
\frac{S}{N}=\left[2T\int_{f_{\rm min}}^{f_{\rm max}}df\,S_h^2(f)\sum_{a=1}^{N_{\rm det}}\sum_{b>a}^{N_{\rm det}}\frac{\Gamma_{ab}^2(f)}{S_n^{(a)}(f)S_n^{(b)}(f)}\right]^{1/2}\,,
\end{equation}
where $\Gamma_{ab}(f)$ is the overlap reduction function (ORF) of the pair $(a,b)$ with $a<b$, specified by the antenna pattern functions of the two detectors $F_a^{A}(\hat{\bf{n}})$, $F_b^{A}(\hat{\bf{n}})$ and their separation $\vec{\bf{x}}_b-\vec{\bf{x}}_a$, as 
\begin{equation}
\Gamma_{ab}(f)=\int_{S^2} \frac{d^2\hat{\bf{n}}}{4\pi}\,\,e^{2\pi i f\hat{\bf{n}}\cdot(\vec{\bf{x}}_b-\vec{\bf{x}}_a)/c}\,\sum_{A\in\{+,\times\}}F_a^{A}(\hat{\bf{n}})F_b^{A}(\hat{\bf{n}})\,.
\end{equation}
Introducing the quantity
\begin{equation}
\Omega_{\rm eff}(f)=\frac{4\pi^2}{3H_0^2}f^3\left[\sum_{a=1}^{N_{\rm det}}\sum_{b>a}^{N_{\rm det}}\frac{\Gamma_{ab}^2(f)}{S_n^{(a)}(f)S_n^{(b)}(f)}\right]^{-1/2}
\end{equation}
and using eq.~(\ref{eq:OmegaGW_Sh}), we get
\begin{equation}
\label{eq: SNR_OmegaGW}
\frac{S}{N}=\left[2T\int_{f_{\rm min}}^{f_{\rm max}}df\,\frac{\Omega^2_{\rm GW}(f)}{\Omega^2_{\rm eff}(f)}\right]^{1/2}\,.
\end{equation}
All the relevant features of the network are encoded in $\Omega_{\rm eff}(f)$.

We now recall how the PLS curve was constructed in~\cite{Thrane:2013oya}.
As the starting point, one considers the family of power-law GW backgrounds\footnote{Another interesting class of GW backgrounds, motivated by early Universe phenomena, is given by broken power laws; see~\cite{Chowdhury:2022gdc, Marriott-Best:2024anh} for the corresponding notion of broken power-law sensitivity (BPLS) curve. However, in this work we only consider the standard power laws, given by eq.~(\ref{eq: pow law GW}).}
\begin{equation}
\label{eq: pow law GW}
\Omega_{\rm GW}(f;\beta)=\Omega_\beta(f/f_{\rm ref})^\beta \,,
\end{equation}
where the exponent $\beta$ is the family parameter which can take any real value $\beta\in\left(-\infty,+\infty\right)$ and $f_{\rm ref}$ is an arbitrarily chosen positive reference frequency. The dependence of the amplitude coefficient $\Omega_\beta$ on $\beta$ is determined by requiring that, in the observation time $T$, the SNR in the frequency band of the network $[f_{\rm min}, f_{\rm max}]$ takes an assigned value $\rho$. Plugging eq.~(\ref{eq: pow law GW}) into eq.~(\ref{eq: SNR_OmegaGW}) gives
\begin{equation}
\label{eq: Omega_beta}
\Omega_\beta=\rho\left[2T\int_{f_{\rm min}}^{f_{\rm max}} d{f}~\Omega^{-2}_{\rm eff}(f)\left(\frac{f}{f_{\rm ref}}\right)^{2\beta}\right]^{-1/2}\,.
\end{equation}

Using eq. (\ref{eq: Omega_beta}), the family of power-law GW backgrounds in eq. (\ref{eq: pow law GW}) is then
\begin{equation}
\label{eq: power law family}
\Omega_{\rm GW}(f;\beta)=\rho\,\frac{\left(\frac{f}{f_{\rm ref}}\right)^{\beta}}{\left[2T\int_{f_{\rm min}}^{f_{\rm max}} d{f'}~\Omega^{-2}_{\rm eff}(f') \left(\frac{f'}{f_{\rm ref}}\right)^{2\beta}\right]^{1/2}}\,,
\end{equation}
where we renamed the dummy integration variable to $f'$, in order to distinguish it from the frequency variable $f$ on which $\Omega_{\rm GW}(f;\beta)$ depends. The choice of the arbitrary $f_{\rm ref}$ plays no role. Indeed, it cancels in eq. (\ref{eq: power law family}) and we only keep it in order to have dimensionless ratios of frequencies as the bases of powers.

\section{The PLS curve}
\label{sec:PLSmain}

\noindent Following the definition given in~\cite{Thrane:2013oya}, ``the envelope of the $\Omega_{\rm gw}(f)$ power-law curves is the power-law integrated sensitivity curve for a correlation measurement using two or more detectors''\footnote{See the beginning of point 5 in section 3A of~\cite{Thrane:2013oya}.}. Adapting it to our notation, the PLS curve is the envelope of the family of power-law backgrounds $\Omega_{\rm GW}(f;\beta)$ in eq.~(\ref{eq: power law family}), where $\beta$ is the family parameter running over all real numbers. It is then immediately claimed in eq.~(30) of~\cite{Thrane:2013oya} that (again, adapting it to our notation\footnote{Eq.~(30) of~\cite{Thrane:2013oya} in its original notation is
\begin{equation}
\Omega_{\rm PI}(f)=\max_\beta\left[\Omega_\beta \left(\frac{f}{f_{\rm ref}}\right)^\beta\right]\,.\nonumber
\end{equation}
})

\begin{equation}
\label{eq: PLS as max}
\Omega_{\rm PLS}(f)=\underset{\beta}{\max}~\Omega_{\rm GW}(f;\beta)\,.
\end{equation}

However, at this stage, this equation cannot be justified on the basis of the primary definition of the PLS as the envelope of a family of backgrounds. In fact, in general an envelope of a family is not even a function, but just a curve, so it is not clear a priori that the resulting envelope takes the form of a function $\Omega_{\rm PLS}(f)$. Furthermore the nature of the stationary point involved cannot be claimed before further investigations, so claiming that we are talking about a maximum as in eq.~(\ref{eq: PLS as max}) is not justified yet. The precise meaning of these observations will become more clear later.
In the following {\it we stick to the primary definition of the PLS curve as an envelope}. As a byproduct of our analysis, we will also {\it prove} in section~\ref{subsec:eq30_explain} and appendix~\ref{app:eq30_details} that, for technical reasons deserving an explanation, eq.~(\ref{eq: PLS as max}) actually holds. However, even if this turns out to work for the problem considered here, in more general situations an identification between the envelope and a function resulting from a maximization procedure, as in eq.~(30) of~\cite{Thrane:2013oya} [i.e. in eq.~(\ref{eq: PLS as max})], does not hold. A relevant case will be discussed in a separate paper~\cite{Belgacem:extPLS}, which will introduce an extended notion of PLS curve useful in the presence of noise correlations between detectors.

\subsection{A remark about the PLS interpretation}
\label{sec:remark1}

It is already possible to give an interpretation of the use of the PLS curve, defined as the envelope of the family~(\ref{eq: power law family}).
Let us consider a given fixed power-law background, for definiteness we name it $\Omega_{\rm GW}(f)=\Omega_0(f/f_{\rm ref})^{\beta_0}$, where $\Omega_0>0$ and $\beta_0$ are two arbitrary real numbers independent from each other and $f_{\rm ref}$ is the same reference value chosen previously. In a log-log plot with $\ln\Omega_{\rm GW}$ on the vertical axis and $\ln f$ on the horizontal one, this GW background looks like a straight line with slope equal to $\beta_0$ (and $\Omega_0$ is its value at $f=f_{\rm ref}$). Let us call $(S/N)_0$ the value for this background computed from eq.~(\ref{eq: SNR_OmegaGW}).

For each real power-law exponent $\beta$ there exists one and only one member of the family~(\ref{eq: power law family}). Since the PLS curve is defined as the envelope of the entire family, in a log-log plot there exists a unique straight line parallel to the specific background that we considered before and tangent to the PLS curve. This unique parallel line is simply obtained by specifying $\beta=\beta_0$ in eq.~(\ref{eq: power law family}). By definition, the SNR of this new background is equal to the value $\rho$ used in the PLS construction. Since the two parallel lines correspond to two backgrounds that only differ by a multiplicative constant, it follows from eq.~(\ref{eq: SNR_OmegaGW}) that the ratio between the SNR of the original background, that we call $(S/N)_0$, and that of the new parallel one, $\rho$, is simply equal to the ratio of their amplitudes:
\begin{equation}
\label{eq:SNRparallel}
\frac{(S/N)_0}{\rho}=\frac{\Omega_0}{\Omega_{\beta_0}}\,,
\end{equation}
where $\Omega_{\beta_0}$ is given by eq.~(\ref{eq: Omega_beta})  with $\beta=\beta_0$.
If the $\rho$ used in the construction of the PLS curve is a threshold value that we decide to take when assessing the detectability of a background, then from eq.~(\ref{eq:SNRparallel}),
\begin{equation}
(S/N)_0\geq\rho\iff\Omega_0\geq\Omega_{\beta_0}\,.
\end{equation}
Therefore we arrive at the following simple interpretation:

{\it A power-law background is detectable if and only if, in a log-log plot, its straight line lies above (or coincides with) the unique straight line parallel to it and tangent to the PLS curve}.

\subsection{Exact parametric equation}
Mathematically, given a family of curves in the $(x,y)$ plane, written as $F(x,y;k)=0$ where $k$ is the family parameter, the envelope of the family is defined as the set of points $(x,y)$ such that, for some $k$, both conditions $F(x,y;k)=0$ and $\frac{\partial}{\partial k}F(x,y;k)=0$ hold. These conditions tell us that the envelope is tangent to each member of the family. As the family parameter $k$ runs over its interval of definition, the envelope is created by all the tangency points. In the case of eq.~(\ref{eq: power law family}), the variables $x$ and $y$ are $f$ and $\Omega_{\rm GW}$, respectively, while the family parameter $k$ is the power-law exponent $\beta$. The family~(\ref{eq: power law family}) can be put into the form $F(f,\Omega_{\rm GW};\beta)=0$ by defining\footnote{There is some freedom in the definition of a $F(f,\Omega_{\rm GW};\beta)$ such that eq.~(\ref{eq: power law family}) is equivalent to $F(f,\Omega_{\rm GW};\beta)=0$. This freedom includes (but is not limited to), for example, the possibility of multiplying $F(f,\Omega_{\rm GW};\beta)$ by an arbitrary non-zero constant. All these choices are legitimate and lead to the same envelope. The one adopted in eq.~(\ref{eq:Fdefinition}) (or its opposite) is the most simple choice.}
\begin{equation}
\label{eq:Fdefinition}
F(f,\Omega_{\rm GW};\beta)\equiv\rho\,\frac{\left(\frac{f}{f_{\rm ref}}\right)^{\beta}}{\left[2T\int_{f_{\rm min}}^{f_{\rm max}} d{f'}~\Omega^{-2}_{\rm eff}(f') \left(\frac{f'}{f_{\rm ref}}\right)^{2\beta}\right]^{1/2}}-\Omega_{\rm GW}\,. 
\end{equation}

Therefore, the envelope of the family~(\ref{eq: power law family}) is defined by the following two equations:
\begin{equation}
\label{eq:F}
F(f,\Omega_{\rm GW};\beta)=\rho\,\frac{\left(\frac{f}{f_{\rm ref}}\right)^{\beta}}{\left[2T\int_{f_{\rm min}}^{f_{\rm max}} d{f'}~\Omega^{-2}_{\rm eff}(f') \left(\frac{f'}{f_{\rm ref}}\right)^{2\beta}\right]^{1/2}}-\Omega_{\rm GW}=0\,   
\end{equation}
and
\begin{eqnarray}
\label{eq:Fderwrtbeta}
\frac{\partial}{\partial \beta}F(f,\Omega_{\rm GW};\beta)&=&\rho\,\frac{\left(\frac{f}{f_{\rm ref}}\right)^{\beta}}{\left[2T\int_{f_{\rm min}}^{f_{\rm max}} d{f'}~\Omega^{-2}_{\rm eff}(f') \left(\frac{f'}{f_{\rm ref}}\right)^{2\beta}\right]^{1/2}}\,\nonumber\\
&&\times\left[\ln\left(\frac{f}{f_{\rm ref}}\right)-\frac{\int_{f_{\rm min}}^{f_{\rm max}} d{f'}~\Omega^{-2}_{\rm eff}\left(f'\right)\left(\frac{f'}{f_{\rm ref}}\right)^{2\beta}\ln{\left(\frac{f'}{f_{\rm ref}}\right)}}{\int_{f_{\rm min}}^{f_{\rm max}} d{f'}~\Omega^{-2}_{\rm eff}\left(f'\right)\left(\frac{f'}{f_{\rm ref}}\right)^{2\beta}}\right]=0\,.\label{eq:derF}
\end{eqnarray}

The solution of eqs.~(\ref{eq:F}) and (\ref{eq:derF}) can be written in parametric form as\footnote{Since the family~(\ref{eq: power law family}) is independent of the choice of $f_{\rm ref}$, the same must be true for the envelope. Indeed, it is straightforward to check that eqs.~(\ref{eq:parfbeta}) and (\ref{eq:parOmPLSbeta}) do not depend on $f_{\rm ref}$.}

\begin{eqnarray}
f(\beta)&=&f_{\rm ref}\,\exp\left[\frac{\int_{f_{\rm min}}^{f_{\rm max}} d{f'}~\Omega^{-2}_{\rm eff}\left(f'\right)\left(\frac{f'}{f_{\rm ref}}\right)^{2\beta}\ln{\left(\frac{f'}{f_{\rm ref}}\right)}}{\int_{f_{\rm min}}^{f_{\rm max}} d{f'}~\Omega^{-2}_{\rm eff}\left(f'\right)\left(\frac{f'}{f_{\rm ref}}\right)^{2\beta}}\right]\,,\label{eq:parfbeta}\\
\Omega_{\rm PLS}(\beta)\equiv\Omega_{\rm GW}(\beta)&=&\rho\,\frac{\exp\left[\beta\frac{\int_{f_{\rm min}}^{f_{\rm max}} d{f'}~\Omega^{-2}_{\rm eff}\left(f'\right)\left(\frac{f'}{f_{\rm ref}}\right)^{2\beta}\ln{\left(\frac{f'}{f_{\rm ref}}\right)}}{\int_{f_{\rm min}}^{f_{\rm max}} d{f'}~\Omega^{-2}_{\rm eff}\left(f'\right)\left(\frac{f'}{f_{\rm ref}}\right)^{2\beta}}\right]}{\left[2T\int_{f_{\rm min}}^{f_{\rm max}} d{f'}~\Omega^{-2}_{\rm eff}(f') \left(\frac{f'}{f_{\rm ref}}\right)^{2\beta}\right]^{1/2}}\,,\label{eq:parOmPLSbeta}
\end{eqnarray}
where in eq.~(\ref{eq:parOmPLSbeta}) we introduced the notation $\Omega_{\rm PLS}(\beta)$ as a reminder that the two equations~(\ref{eq:parfbeta}) and (\ref{eq:parOmPLSbeta}) describe the PLS curve. More precisely, as the parameter $\beta$ runs over all real numbers, the point of coordinates $(f(\beta),\Omega_{\rm PLS}(\beta))$ describes a curve. This curve, being the envelope of the family~(\ref{eq: power law family}), is, by definition, the PLS curve. Since eqs.~(\ref{eq:parfbeta}) and (\ref{eq:parOmPLSbeta}) are exact, they give the most precise and efficient way to compute the PLS curve, rather then going numerically through the maximization problem in eq.~(\ref{eq: PLS as max}), which is not even yet justified at this stage.

\section{The PLS curve is a function}
\label{sec:PLSisfunc}
If we want to be able to write the PLS curve as a function of frequency $\Omega_{\rm PLS}(f)$ [as the left-hand side of eq.~(\ref{eq: PLS as max}), repeating eq.~(30) of~\cite{Thrane:2013oya}, does], then we have to show that $f(\beta)$ in eq.~(\ref{eq:parfbeta}) is an invertible function. Only in this case it is true that, for each f, there is a unique corresponding $\beta(f)$ and then eq.~(\ref{eq:parOmPLSbeta}) delivers a unique result for that $f$, given by $\Omega_{\rm PLS}(\beta(f))$ [which, for brevity, we can then call $\Omega_{\rm PLS}(f)$].

It is convenient to introduce the following quantity:
\begin{equation}
\label{eq:pdist}
p(f;\beta)\equiv\frac{\Omega^{-2}_{\rm eff}\left(f\right)\left(\frac{f}{f_{\rm ref}}\right)^{2\beta}}{\int_{f_{\rm min}}^{f_{\rm max}}d{f'}~\Omega^{-2}_{\rm eff}\left(f'\right)\left(\frac{f'}{f_{\rm ref}}\right)^{2\beta}}\,.
\end{equation}
In terms of $p(f;\beta)$, eq.~(\ref{eq:parfbeta}), can be simply recast into
\begin{equation}
\label{eq:fbetawithp}
\ln\left[\frac{f(\beta)}{f_{\rm ref}}\right]=\int_{f_{\rm min}}^{f_{\rm max}}df'~p(f';\beta)\ln{\left(\frac{f'}{f_{\rm ref}}\right)}\,,
\end{equation}
where again the choice of $f_{\rm ref}$ is irrelevant.
Note that
\begin{equation}
\label{eq:pdistdef}
p(f;\beta)>0\,\qquad{\rm and}\qquad \int_{f_{\rm min}}^{f_{\rm max}}df~p(f;\beta)=1\,.
\end{equation}
Therefore eq.~(\ref{eq:fbetawithp}) is saying that $\ln\left[f(\beta)/f_{\rm ref}\right]$ can be seen as the average value of $\ln\left(f'/f_{\rm ref}\right)$ over the positive and normalized $\beta$-dependent distribution $p(f';\beta)$.
To show the invertibility of $f(\beta)$ we take the derivative with respect to $\beta$ of eq.~(\ref{eq:fbetawithp}) using also the definition~(\ref{eq:pdist}). After some calculations, the result is
\begin{equation}
\label{eq:der_fbetawithp}
\frac{1}{f(\beta)}\frac{df(\beta)}{d\beta}=2\left\{\int_{f_{\rm min}}^{f_{\rm max}}d{f'}~p(f';\beta)\left[\ln{\left(\frac{f'}{f_{\rm ref}}\right)}\right]^2-\left[\int_{f_{\rm min}}^{f_{\rm max}}df'~p(f';\beta)\ln{\left(\frac{f'}{f_{\rm ref}}\right)}\right]^2\right\}\,.
\end{equation}
Apart from the overall factor of 2, the right-hand side of eq. (\ref{eq:der_fbetawithp}) is nothing but the variance of $\ln{\left(f'/f_{\rm ref}\right)}$ over the positive and normalized distribution $p(f';\beta)$.
Therefore it is positive, as also evident from its equivalent form obtained using eqs.~(\ref{eq:fbetawithp}) and (\ref{eq:pdistdef}):
\begin{equation}
\label{eq:der_fbetawithp_equiv}
\frac{1}{f(\beta)}\frac{df(\beta)}{d\beta}=2\int_{f_{\rm min}}^{f_{\rm max}}d{f'}~p(f';\beta)\left[\ln{\left(\frac{f'}{f(\beta)}\right)}\right]^2>0\,.
\end{equation}
Since the sign of the result in eq.~(\ref{eq:der_fbetawithp_equiv}) does not change, we deduce that $f(\beta)$ is injective. The sign is positive, which also specifies that $f(\beta)$ is strictly increasing.

We can also observe that, when $\beta\to-\infty$, the distribution $p(f;\beta)$ in eq.~(\ref{eq:pdist}), which is normalized in the band $[f_{\rm min}, f_{\rm max}]$ (see eq.~(\ref{eq:pdistdef})), is dominated by the lowest frequency, i.e. by $f_{\rm min}$. More precisely, for any $f$ with $f_{\rm min}<f\leq f_{\rm max}$,
\begin{equation}
\lim_{\beta\to-\infty}\frac{p(f;\beta)}{p(f_{\rm min};\beta)}=0\,.
\end{equation}
Then, when $\beta\to-\infty$,  the right-hand side of eq.~(\ref{eq:fbetawithp}) approaches $\ln{(f_{\rm min}/f_{\rm ref})}$ and eq.~(\ref{eq:fbetawithp}) gives
\begin{equation}
\label{eq:limfmin}
\lim_{\beta\to-\infty}f(\beta)=f_{\rm min}\,.
\end{equation}
Similarly, when $\beta\to+\infty$,
\begin{equation}
\label{eq:limfmax}
\lim_{\beta\to+\infty}f(\beta)=f_{\rm max}\,.
\end{equation}
Eqs.~(\ref{eq:limfmin}) and (\ref{eq:limfmax}), together with the strict monotonicity of $f(\beta)$, mean that when $\beta$ runs over the real numbers, $f(\beta)$ runs strictly monotonically over the open\footnote{The bounds $f_{\rm min}$ and $f_{\rm max}$ are excluded because they are only obtained in the limits $\beta\to-\infty$ and $\beta\to+\infty$, respectively.} interval $(f_{\rm min}, f_{\rm max})$. Thus we conclude that $f(\beta)$ is invertible, with the domain of the inverse function $\beta(f)$ being the open interval $(f_{\rm min}, f_{\rm max})$.
Therefore, following the argument presented at the beginning of this section, we have proved that the PLS curve is a function of frequency, that we call $\Omega_{\rm PLS}(f)$. Notice that, since the function $f(\beta)$ is strictly increasing, the inverse function $\beta(f)$ is strictly increasing too.

\subsection{The PLS function as a maximization problem}
\label{subsec:eq30_explain}
We can now briefly discuss why eq.~(30) of~\cite{Thrane:2013oya} [that we replicated in eq.~(\ref{eq: PLS as max})] turns out to be equivalent to the primary definition of the PLS curve as the envelope of a family of power-law backgrounds. First of all, we explained in section \ref{sec:PLSisfunc} why the PLS {\it curve}, i.e. the set of points $(f(\beta),\Omega_{\rm PLS}(\beta))$ with $\beta\in\mathbb{R}$ in eqs.~(\ref{eq:parfbeta}) and (\ref{eq:parOmPLSbeta}), is also a {\it function} $\Omega_{\rm PLS}(f)$. Note that this was not clear a priori and, indeed, the proof required an explicit computation to show the invertibility of $f(\beta)$. Now that the left-hand side of eq.~(\ref{eq: PLS as max}) makes sense, the missing piece is an explanation of the right-hand side. This is done in detail in appendix~\ref{app:eq30_details} and it basically amounts to viewing the same computations done so far from a slightly different perspective. Incidentally, the discussion in appendix~\ref{app:eq30_details} also guarantees that, if one keeps using eq.~(30) of~\cite{Thrane:2013oya} to compute the PLS function, then, for each frequency $f$, a maximization algorithm can only converge to a unique value (that is the global maximum that we are looking for), which is good. However, it is not convenient to use such an approach, because the parametric equations~(\ref{eq:parfbeta}), (\ref{eq:parOmPLSbeta}) already give exact points of the PLS function and therefore they are more precise and efficient than going through a maximization algorithm for each $f$.

\section{Shape of the PLS function and peak sensitivity}
\label{sec:PLSshape}
Let us now derive some other properties of the PLS function $\Omega_{\rm PLS}(f)$. By construction, we expect that, in a log-log plot, the slope of the PLS function at a frequency $f$ is given by $\beta(f)$ [where $\beta(f)$ is the inverse function of $f(\beta)$ in eq.~(\ref{eq:parfbeta}); the inverse function exists, as shown in section~\ref{sec:PLSisfunc}]. As a check of eqs.~(\ref{eq:parfbeta}) and (\ref{eq:parOmPLSbeta}), one can also confirm this expectation by an explicit calculation. A convenient way to to achieve the result is the following. First, using eq.~(\ref{eq:parfbeta}), we write eq.~(\ref{eq:parOmPLSbeta}) in the precursor form
\begin{equation}
\label{eq:PLSparamprecursor}
\Omega_{\rm PLS}(\beta)=\rho\,\frac{\left(\frac{f(\beta)}{f_{\rm ref}}\right)^\beta}{\left[2T\int_{f_{\rm min}}^{f_{\rm max}} d{f'}~\Omega^{-2}_{\rm eff}(f') \left(\frac{f'}{f_{\rm ref}}\right)^{2\beta}\right]^{1/2}}\,.
\end{equation}
Taking the logarithmic derivative with respect to $\beta$ of eq.~(\ref{eq:PLSparamprecursor}), we get
\begin{eqnarray}
\frac{d\ln{\Omega_{\rm PLS}(\beta)}}{d\beta}&=&\ln\left(\frac{f(\beta)}{f_{\rm ref}}\right)+\beta\,\frac{d\ln{f(\beta)}}{d\beta}-\frac{\int_{f_{\rm min}}^{f_{\rm max}} d{f'}~\Omega^{-2}_{\rm eff}\left(f'\right)\left(\frac{f'}{f_{\rm ref}}\right)^{2\beta}\ln{\left(\frac{f'}{f_{\rm ref}}\right)}}{\int_{f_{\rm min}}^{f_{\rm max}} d{f'}~\Omega^{-2}_{\rm eff}\left(f'\right)\left(\frac{f'}{f_{\rm ref}}\right)^{2\beta}}\,\nonumber\\
&=&\beta\,\frac{d\ln{f(\beta)}}{d\beta}\,,
\end{eqnarray}
where, to obtain the last equality, we observed that the first and third addends cancel because of eq.~(\ref{eq:parfbeta}).
Therefore the slope of $\Omega_{\rm PLS}(f)$ in a log-log plot is
\begin{equation}
\label{eq: loglog slope PLS}
\frac{d\ln{\Omega_{\rm PLS}(f)}}{d\ln{f}}=\left[\frac{\frac{d\ln{\Omega_{\rm PLS}(\beta)}}{d\beta}}{\frac{d\ln{f(\beta)}}{d\beta}}\right]_{\beta=\beta(f)}=\beta(f)\,,
\end{equation}
as expected.

A first qualitative description of the PLS function $\Omega_{\rm PLS}(f)$ in a log-log plot can be extracted from eq.~(\ref{eq: loglog slope PLS}), remembering also that $f(\beta)$ in eq.~(\ref{eq:parfbeta}) is a strictly increasing function.
When the parameter $\beta$ runs from $-\infty$ to $0$, the frequency $f(\beta)$ increases and the PLS function decreases slower and slower. When $\beta$ goes from $0$ to $+\infty$, the frequency $f(\beta)$ increases and the PLS function grows faster and faster.
Therefore $\beta=0$ gives the unique minimum (relative and absolute) of the PLS function. Let us denote by $f_{\rm peak}\equiv f(\beta=0)$ and $\Omega_{\rm peak}\equiv \Omega_{\rm PLS}(\beta=0)$ the corresponding values. The peak sensitivity\footnote{This terminology was used e.g. in appendix B of~\cite{Branchesi:2023mws}.} $\Omega_{\rm peak}$ can be readily obtained from eq.~(\ref{eq:parOmPLSbeta}) with $\beta=0$:

\begin{equation}
\label{eq: peak sensitivity result}
\Omega_{\rm peak}=\rho\left[2T\int_{f_{\rm min}}^{f_{\rm max}} d{f}~\Omega^{-2}_{\rm eff}\left(f\right)\right]^{-1/2}\,,
\end{equation}
where we renamed the dummy integration variable $f'$ to $f$ since there is no ambiguity in eq.~(\ref{eq: peak sensitivity result}). This is an exact and ready-to-use formula for the peak sensitivity. For example, it can be used to improve (both in time and precision) the computation of the peak sensitivity of network depending on some internal parameter entering $\Omega_{\rm eff}$ (e.g. for plots like Fig.~71 of~\cite{Branchesi:2023mws}).\footnote{Eq.~(\ref{eq: peak sensitivity result}) simply requires one integration for each configuration, i.e. for a given $\Omega_{\rm eff}(f)$. This is significantly more efficient than determining the full PLS function and then searching numerically for its minimum. The procedure slows down even further if the PLS function is evaluated by computing eq. (30) of~\cite{Thrane:2013oya} [i.e. eq.~(\ref{eq: PLS as max})] with some maximization algorithm, rather than using the exact parametric equation~(\ref{eq:parfbeta}), (\ref{eq:parOmPLSbeta}).} Similarly we also get an exact equation for $f_{\rm peak}$ by plugging $\beta=0$ in eq.~(\ref{eq:parfbeta}):
\begin{equation}
\label{eq:fpeakresult}
f_{\rm peak}=f_{\rm ref}\,\exp\left[\frac{\int_{f_{\rm min}}^{f_{\rm max}} d{f}~\Omega^{-2}_{\rm eff}\left(f\right)\ln{\left(\frac{f}{f_{\rm ref}}\right)}}{\int_{f_{\rm min}}^{f_{\rm max}} d{f}~\Omega^{-2}_{\rm eff}\left(f\right)}\right]\,,
\end{equation}
which, as always, is not affected by the choice of the arbitrary $f_{\rm ref}$.

As a final observation, taking a derivative of eq.~(\ref{eq: loglog slope PLS}) with respect to $\ln{f}$, one gets
\begin{equation}
\label{eq: loglog convexity PLS}
\frac{{\rm d^2}\ln{\Omega_{\rm PLS}(f)}}{{\rm d}(\ln{f})^2}=f\,\frac{d\beta(f)}{d{f}}>0\,,
\end{equation}
where, for the inequality, we used that $\beta(f)$ is a strictly increasing function, as shown in section~\ref{sec:PLSisfunc}. We conclude from eq.~(\ref{eq: loglog convexity PLS}) that {\it in a log-log plot the PLS function is strictly convex}.

Comparing eq.~(\ref{eq: loglog convexity PLS}) with eq.~(\ref{eq:der_fbetawithp_equiv}), we can also write
\begin{eqnarray}
\frac{{\rm d^2}\ln{\Omega_{\rm PLS}(f)}}{{\rm d}(\ln{f})^2}&=&\frac{1}{2\int_{f_{\rm min}}^{f_{\rm max}}d{f'}~p(f';\beta(f))\left[\ln{\left(\frac{f'}{f}\right)}\right]^2}\,\nonumber\\
&=&\frac{1}{2}\frac{\int_{f_{\rm min}}^{f_{\rm max}}d{f'}~\Omega^{-2}_{\rm eff}\left(f'\right)\left(\frac{f'}{f_{\rm ref}}\right)^{2\beta(f)}}{\int_{f_{\rm min}}^{f_{\rm max}}d{f'}~\Omega^{-2}_{\rm eff}\left(f'\right)\left(\frac{f'}{f_{\rm ref}}\right)^{2\beta(f)}\left[\ln{\left(\frac{f'}{f}\right)}\right]^2}\,,
\end{eqnarray}
where, in the last equality, we used eq.~(\ref{eq:pdist}).

\section{Conclusions and outlook}
\label{sec:concl}
We revisited critically the notion of power-law sensitivity curve and derived some exact results for it. The purpose of the paper is both conceptual and practical, with the two aspects merging along the derivations. Among the conceptual points, we discussed how the primary definition of the PLS curve as an envelope of a family of power-law backgrounds, which admits a simple interpretation, is equivalent to eq.~(30) of~\cite{Thrane:2013oya}. On the practical side, the analysis of the envelope led us to an exact parametric equation for the PLS curve, see eqs.~(\ref{eq:parfbeta}) and (\ref{eq:parOmPLSbeta}). This provides a more efficient and precise method of computation, compared to a numerical maximization procedure. We also derived some general features of the PLS shape, including convexity and an exact result for the peak sensitivity (and the frequency where it is reached), see eqs.~(\ref{eq: peak sensitivity result}) and (\ref{eq:fpeakresult}). Again, this is also relevant for numerical computations as it reduces significantly the number of steps needed. The analytical methods developed in this paper will be applied to the more difficult case of noise correlations in~\cite{Belgacem:extPLS}, where some extra care will be necessary when dealing with a new notion of extended PLS curve.

\begin{acknowledgments}
The author is supported by the SNSF grant CRSII5$\_$213497. We thank Michele Maggiore for useful comments. We also thank Francesco Iacovelli and Niccol\`o Muttoni for numerical tests.
\end{acknowledgments}

\appendix
\section{Details on section~\ref{subsec:eq30_explain}}
\label{app:eq30_details}

The right-hand side of eq.~(\ref{eq: PLS as max}) (taken from eq.~(30) of~\cite{Thrane:2013oya}) is a maximization problem over $\beta$, for each frequency $f$. More precisely, for a given frequency $f$, we want to find the value $\bar{\beta}(f)$ of $\beta$ at which $\Omega_{\rm GW}(f;\beta)$ reaches its global maximum.
Let us start by finding stationary points $\bar{\beta}(f)$ of the function $\Omega_{\rm GW}(f;\beta)$ in eq.~(\ref{eq: power law family}) for a given $f$. Then the following equation must hold at $\beta=\bar{\beta}(f)$:
\begin{equation}
\label{eq: beta bar condition}
\frac{\partial}{\partial\beta}\Omega_{\rm GW}(f;\beta)\bigg|_{\beta=\bar{\beta}(f)}=0\,.
\end{equation}
At this stage $\bar{\beta}(f)$ is a generic stationary point, but at the end of the appendix we will reach the conclusion that it is unique and that it really is the global maximum.

One can derive
\begin{equation}
\label{eq: der beta}
\frac{\partial}{\partial\beta}\Omega_{\rm GW}(f;\beta)=\Omega_{\rm GW}(f;\beta)\left[\ln{\left(\frac{f}{f_{\rm ref}}\right)}-\frac{\int_{f_{\rm min}}^{f_{\rm max}} d{f'}~\Omega^{-2}_{\rm eff}\left(f'\right)\left(\frac{f'}{f_{\rm ref}}\right)^{2\beta}\ln{\left(\frac{f'}{f_{\rm ref}}\right)}}{\int_{f_{\rm min}}^{f_{\rm max}} d{f'}~\Omega^{-2}_{\rm eff}\left(f'\right)\left(\frac{f'}{f_{\rm ref}}\right)^{2\beta}}\right]\,,
\end{equation}
where again the arbitrary reference frequency $f_{\rm ref}$ is irrelevant. Eq.~(\ref{eq: der beta}) is the same as eq.~(\ref{eq:derF}) with $\Omega_{\rm GW}(f;\beta)$ given in eq.~(\ref{eq: power law family}). Of course, this is not accidental: it is simply due to the fact that, by construction, the quantity $F(f,\Omega_{\rm GW}; \beta)$ in eq.~(\ref{eq:F}) is just $F(f,\Omega_{\rm GW}; \beta)=\Omega_{\rm GW}(f;\beta)-\Omega_{\rm GW}$, with $\Omega_{\rm GW}(f;\beta)$ given in eq.~(\ref{eq: power law family}), hence $\frac{\partial}{\partial\beta}F(f,\Omega_{\rm GW}; \beta)=\frac{\partial}{\partial\beta}\Omega_{\rm GW}(f;\beta)$.

The stationarity condition (\ref{eq: beta bar condition}) is therefore the same as eq.~(\ref{eq:parfbeta}) with $\beta=\bar{\beta}(f)$, that is
\begin{equation}
\label{eq: beta bar equation}
\frac{\int_{f_{\rm min}}^{f_{\rm max}} d{f'}~\Omega^{-2}_{\rm eff}\left(f'\right)\left(\frac{f'}{f_{\rm ref}}\right)^{2\bar{\beta}(f)}\ln{\left(\frac{f'}{f_{\rm ref}}\right)}}{\int_{f_{\rm min}}^{f_{\rm max}} d{f'}~\Omega^{-2}_{\rm eff}\left(f'\right)\left(\frac{f'}{f_{\rm ref}}\right)^{2\bar{\beta}(f)}}=\ln{\left(\frac{f}{f_{\rm ref}}\right)}\,.
\end{equation}
Thus the arguments in section~\ref{sec:PLSisfunc} already imply that, for every $f$ with $f_{\rm min}<f<f_{\rm max}$, a solution $\bar{\beta}(f)$ to eq.~(\ref{eq: beta bar equation}) exists and is unique. So far, we have shown existence and uniqueness of the stationary point $\bar{\beta}(f)$ satisfying eq.~(\ref{eq: beta bar condition}). A closer look also reveals the nature of the stationary point, as follows. For convenience, let us give a name to the function of $\beta$ appearing on the right-hand side of eq.~(\ref{eq:fbetawithp}), introducing
\begin{equation}
\label{eq: F def}
A(\beta)\equiv\int_{f_{\rm min}}^{f_{\rm max}}d{f'}~p(f';\beta)\ln{\left(\frac{f'}{f_{\rm ref}}\right)}\,.
\end{equation}
We rewrite eq.~(\ref{eq: der beta}) as
\begin{equation}
\label{eq: der OmegaGW with A}
\frac{\partial}{\partial\beta}\ln{\Omega_{\rm GW}(f;\beta)}=\ln{\left(\frac{f}{f_{\rm ref}}\right)}-A(\beta)\,
\end{equation}
and eq.~(\ref{eq: beta bar equation}) as
\begin{equation}
\label{eq:Abetabarf}
A(\bar{\beta}(f))=\ln{\left(\frac{f}{f_{\rm ref}}\right)}\,.
\end{equation}
For a given frequency $f$, let us study the right-hand side of eq.~(\ref{eq: der OmegaGW with A}) as a function of $\beta$. We know that it vanishes for $\beta=\bar{\beta}(f)$. Furthermore the same computations done in eqs.~(\ref{eq:der_fbetawithp}) and (\ref{eq:der_fbetawithp_equiv}) also tell us that $\frac{dA(\beta)}{d\beta}>0$, i.e. $A(\beta)$ is a strictly increasing function. Thus, for all $\beta<\bar{\beta}(f)$, we can be sure that $A(\beta)<A(\bar{\beta}(f))=\ln{\left(\frac{f}{f_{\rm ref}}\right)}$, where in the last equality we used eq.~(\ref{eq:Abetabarf}). Then, from eq.~(\ref{eq: der OmegaGW with A}), we see that for all $\beta<\bar{\beta}(f)$ the inequality $\frac{\partial}{\partial\beta}\ln{\Omega_{\rm GW}(f;\beta)}>0$ holds. Similarly, one can show that for all $\beta>\bar{\beta}(f)$ the inequality $\frac{\partial}{\partial\beta}\ln{\Omega_{\rm GW}(f;\beta)}<0$ holds. These two statements, together with eq.~(\ref{eq: beta bar condition}) imply that $\beta=\bar{\beta}(f)$ is a relative maximum of $\ln{\Omega_{\rm GW}(f;\beta)}$ (for a fixed $f$, as a function of $\beta$) and, equivalently, a relative maximum of $\Omega_{\rm GW}(f;\beta)$. But since we also proved the {\it uniqueness} of the stationary point $\bar{\beta}(f)$, then it must also be the global maximum of $\ln{\Omega_{\rm GW}(f;\beta)}$ and, equivalently, of $\Omega_{\rm GW}(f;\beta)$.
Finally, we can compute this maximum value of $\Omega_{\rm GW}(f;\beta)$ by substituting $\beta=\bar{\beta}(f)$ into eq.~(\ref{eq: power law family}):
\begin{equation}
\label{eq:maxOmbeta}
\max_\beta\Omega_{\rm GW}(f;\beta)=\Omega_{\rm GW}(f;\bar{\beta}(f))=\rho\,\frac{\left(\frac{f}{f_{\rm ref}}\right)^{\bar{\beta}(f)}}{\left[2T\int_{f_{\rm min}}^{f_{\rm max}} d{f'}~\Omega^{-2}_{\rm eff}(f') \left(\frac{f'}{f_{\rm ref}}\right)^{2\bar{\beta}(f)}\right]^{1/2}}\,,
\end{equation}
or, using eq.~(\ref{eq: beta bar equation}),
\begin{equation}
\label{eq:maxOmbeta_fin}
\max_\beta\Omega_{\rm GW}(f;\beta)=\Omega_{\rm GW}(f;\bar{\beta}(f))=\rho\,\frac{\exp\left[\bar{\beta}(f)\,\frac{\int_{f_{\rm min}}^{f_{\rm max}} d{f'}~\Omega^{-2}_{\rm eff}\left(f'\right)\left(\frac{f'}{f_{\rm ref}}\right)^{2\bar{\beta}(f)}\ln{\left(\frac{f'}{f_{\rm ref}}\right)}}{\int_{f_{\rm min}}^{f_{\rm max}} d{f'}~\Omega^{-2}_{\rm eff}\left(f'\right)\left(\frac{f'}{f_{\rm ref}}\right)^{2\bar{\beta}(f)}}\right]}{\left[2T\int_{f_{\rm min}}^{f_{\rm max}} d{f'}~\Omega^{-2}_{\rm eff}(f') \left(\frac{f'}{f_{\rm ref}}\right)^{2\bar{\beta}(f)}\right]^{1/2}}\,.
\end{equation}
As we have already seen, eq.~(\ref{eq: beta bar equation}) is the same as eq.~(\ref{eq:parfbeta}) with $\beta=\bar{\beta}(f)$. In other words, the quantity that we called $\bar{\beta}(f)$ in this appendix is exactly the inverse function of the $f(\beta)$ in eq.~(\ref{eq:parfbeta}) [i.e. the inverse function that we simply called $\beta(f)$ in section~\ref{sec:PLSisfunc}]. Keeping this in mind, we can also see that eq.~(\ref{eq:maxOmbeta_fin}) is the same as eq.~(\ref{eq:parOmPLSbeta}).

These observations complete the proof that eq.~(30) of~\cite{Thrane:2013oya}, i.e. eq.~(\ref{eq: PLS as max}), is true.

\bibliographystyle{utphys}
\bibliography{pls}

\providecommand{\href}[2]{#2}\begingroup\raggedright\begin{thebibliography}{10}

\bibitem{LIGOScientific:2016aoc}
{\bfseries LIGO Scientific, Virgo} Collaboration, B.~P. Abbott {\em et~al.}, ``{Observation of Gravitational Waves from a Binary Black Hole Merger},'' \href{http://dx.doi.org/10.1103/PhysRevLett.116.061102}{{\em Phys. Rev. Lett.} {\bfseries 116} no.~6, (2016) 061102}, \href{http://arxiv.org/abs/1602.03837}{{\ttfamily arXiv:1602.03837 [gr-qc]}}.

\bibitem{Belgacem:2024ohp}
E.~Belgacem, F.~Iacovelli, M.~Maggiore, M.~Mancarella, and N.~Muttoni, ``{The spectral density of astrophysical stochastic backgrounds},''  (11, 2024) , \href{http://arxiv.org/abs/2411.04028}{{\ttfamily arXiv:2411.04028 [gr-qc]}}.

\bibitem{Maggiore:2018sht}
M.~Maggiore, {\em {Gravitational Waves. Vol. 2: Astrophysics and Cosmology}}.
\newblock Oxford University Press,
2018.
\newblock

\bibitem{Christensen:2018iqi}
N.~Christensen, ``{Stochastic Gravitational Wave Backgrounds},'' \href{http://dx.doi.org/10.1088/1361-6633/aae6b5}{{\em Rept. Prog. Phys.} {\bfseries 82} no.~1, (2019) 016903}, \href{http://arxiv.org/abs/1811.08797}{{\ttfamily arXiv:1811.08797 [gr-qc]}}.

\bibitem{NANOGrav:2023gor}
{\bfseries NANOGrav} Collaboration, G.~Agazie {\em et~al.}, ``{The NANOGrav 15 yr Data Set: Evidence for a Gravitational-wave Background},'' \href{http://dx.doi.org/10.3847/2041-8213/acdac6}{{\em Astrophys. J. Lett.} {\bfseries 951} no.~1, (2023) L8}, \href{http://arxiv.org/abs/2306.16213}{{\ttfamily arXiv:2306.16213 [astro-ph.HE]}}.

\bibitem{Hild:2008ng}
S.~Hild, S.~Chelkowski, and A.~Freise, ``{Pushing towards the ET sensitivity using 'conventional' technology},''
\href{http://arxiv.org/abs/0810.0604}{{\ttfamily arXiv:0810.0604 [gr-qc]}}.

\bibitem{Punturo:2010zz}
M.~Punturo {\em et~al.}, ``{The Einstein Telescope: A third-generation gravitational wave observatory},''
\href{http://dx.doi.org/10.1088/0264-9381/27/19/194002}{{\em Class. Quant. Grav.} {\bfseries 27} (2010) 194002}.

\bibitem{Hild:2010id}
S.~Hild {\em et~al.}, ``{Sensitivity Studies for Third-Generation Gravitational Wave Observatories},'' \href{http://dx.doi.org/10.1088/0264-9381/28/9/094013}{{\em Class. Quant. Grav.} {\bfseries 28} (2011) 094013},
\href{http://arxiv.org/abs/1012.0908}{{\ttfamily arXiv:1012.0908 [gr-qc]}}.

\bibitem{ET:2019dnz}
{\bfseries ET} Collaboration, M.~Maggiore {\em et~al.}, ``{Science Case for the Einstein Telescope},'' \href{http://dx.doi.org/10.1088/1475-7516/2020/03/050}{{\em JCAP} {\bfseries 03} (2020) 050}, \href{http://arxiv.org/abs/1912.02622}{{\ttfamily arXiv:1912.02622 [astro-ph.CO]}}.

\bibitem{Branchesi:2023mws}
M.~Branchesi {\em et~al.}, ``{Science with the Einstein Telescope: a comparison of different designs},'' \href{http://dx.doi.org/10.1088/1475-7516/2023/07/068}{{\em JCAP} {\bfseries 07} (2023) 068}, \href{http://arxiv.org/abs/2303.15923}{{\ttfamily arXiv:2303.15923 [gr-qc]}}.

\bibitem{Reitze:2019iox}
D.~Reitze {\em et~al.}, ``{Cosmic Explorer: The U.S. Contribution to Gravitational-Wave Astronomy beyond LIGO},'' {\em Bull. Am. Astron. Soc.} {\bfseries 51} no.~7, (2019) 035, \href{http://arxiv.org/abs/1907.04833}{{\ttfamily arXiv:1907.04833 [astro-ph.IM]}}.

\bibitem{Evans:2021gyd}
M.~Evans {\em et~al.}, ``{A Horizon Study for Cosmic Explorer: Science, Observatories, and Community},'' \href{http://arxiv.org/abs/2109.09882}{{\ttfamily arXiv:2109.09882 [astro-ph.IM]}}.

\bibitem{Evans:2023euw}
M.~Evans {\em et~al.}, ``{Cosmic Explorer: A Submission to the NSF MPSAC ngGW Subcommittee},'' \href{http://arxiv.org/abs/2306.13745}{{\ttfamily arXiv:2306.13745 [astro-ph.IM]}}.

\bibitem{Belgacem:2024ntv}
E.~Belgacem, F.~Iacovelli, M.~Maggiore, M.~Mancarella, and N.~Muttoni, ``{Confusion noise from astrophysical backgrounds at third-generation gravitational-wave detector networks},''  (11, 2024) , \href{http://arxiv.org/abs/2411.04029}{{\ttfamily arXiv:2411.04029 [gr-qc]}}.

\bibitem{Thrane:2013oya}
E.~Thrane and J.~D. Romano, ``{Sensitivity curves for searches for gravitational-wave backgrounds},'' \href{http://dx.doi.org/10.1103/PhysRevD.88.124032}{{\em Phys. Rev. D} {\bfseries 88} no.~12, (2013) 124032}, \href{http://arxiv.org/abs/1310.5300}{{\ttfamily arXiv:1310.5300 [astro-ph.IM]}}.

\bibitem{Maggiore:2007ulw}
M.~Maggiore, {\em {Gravitational Waves. Vol. 1: Theory and Experiments}}.
\newblock Oxford University Press, 2007.

\bibitem{Allen:1997ad}
B.~Allen and J.~D. Romano, ``{Detecting a stochastic background of gravitational radiation: Signal processing strategies and sensitivities},'' \href{http://dx.doi.org/10.1103/PhysRevD.59.102001}{{\em Phys. Rev. D} {\bfseries 59} (1999) 102001}, \href{http://arxiv.org/abs/gr-qc/9710117}{{\ttfamily arXiv:gr-qc/9710117}}.

\bibitem{Chowdhury:2022gdc}
D.~Chowdhury, G.~Tasinato, and I.~Zavala, ``{The rise of the primordial tensor spectrum from an early scalar-tensor epoch},'' \href{http://dx.doi.org/10.1088/1475-7516/2022/08/010}{{\em JCAP} {\bfseries 08} no.~08, (2022) 010}, \href{http://arxiv.org/abs/2204.10218}{{\ttfamily arXiv:2204.10218 [gr-qc]}}.

\bibitem{Marriott-Best:2024anh}
A.~Marriott-Best, D.~Chowdhury, A.~Ghoshal, and G.~Tasinato, ``{Exploring cosmological gravitational wave backgrounds through the synergy of LISA and ET},'' \href{http://arxiv.org/abs/2409.02886}{{\ttfamily arXiv:2409.02886 [astro-ph.CO]}}.

\bibitem{Belgacem:extPLS}
E.~Belgacem, ``{Extended power-law sensitivity curve in the presence of noise correlations},'' {\em in preparation} .

\end{thebibliography}\endgroup

\end{document}